\documentclass[conference]{IEEEtran}
\IEEEoverridecommandlockouts

\usepackage{cite}
\usepackage{amsmath,amssymb,amsfonts}
\usepackage{algorithmic}
\usepackage{graphicx}
\usepackage{textcomp}
\usepackage{xcolor}
\usepackage{url}
\usepackage{booktabs}
\usepackage{subfig}

\def\BibTeX{{\rm B\kern-.05em{\sc i\kern-.025em b}\kern-.08em
    T\kern-.1667em\lower.7ex\hbox{E}\kern-.125emX}}
\begin{document}

\title{Global-local Fourier Neural Operator for Accelerating Coronal Magnetic Field Model}

\author{\IEEEauthorblockN{Yutao Du}
\IEEEauthorblockA{\textit{Mechanical and Industrial Engineering} \\
\textit{NJIT}, Newark, USA \\
yd288@njit.edu} \\
\IEEEauthorblockN{Mengnan Du}
\IEEEauthorblockA{\textit{Department of Data Science} \\
\textit{NJIT}, Newark, USA \\
mengnan.du@njit.edu}

\and
\IEEEauthorblockN{Qin Li}
\IEEEauthorblockA{\textit{Department of Physics} \\
\textit{NJIT}, Newark, USA \\
ql47@njit.edu}\\
\IEEEauthorblockN{Haimin Wang}
\IEEEauthorblockA{\textit{Department of Physics} \\
\textit{NJIT}, Newark, USA \\
haimin.wang@njit.edu}

\and
\IEEEauthorblockN{Raghav Gnanasambandam}
\IEEEauthorblockA{\textit{Industrial and Manufacturing Engineering} \\
\textit{FAMU-FSU}, Tallahassee, FL \\
raghavg@eng.famu.fsu.edu}\\

\IEEEauthorblockN{Bo Shen*\thanks{*Dr. Bo Shen is the corresponding author.}}
\IEEEauthorblockA{\textit{Mechanical and Industrial Engineering} \\
\textit{NJIT}, Newark, USA \\
bo.shen@njit.edu}
}

\maketitle

\allowdisplaybreaks

\begin{abstract}
Exploring the outer atmosphere of the sun has remained a significant bottleneck in astrophysics, given the intricate magnetic formations that significantly influence diverse solar events. Magnetohydrodynamics (MHD) simulations allow us to model the complex interactions between the sun's plasma, magnetic fields, and the surrounding environment. However, MHD simulation is extremely time-consuming, taking days or weeks for simulation. The goal of this study is to accelerate coronal magnetic field simulation using deep learning, specifically, the Fourier Neural Operator (FNO). FNO has been proven to be an ideal tool for scientific computing and discovery in the literature. In this paper, we proposed a global-local Fourier Neural Operator (GL-FNO) that contains two branches of FNOs: the global FNO branch takes downsampled input to reconstruct global features while the local FNO branch takes original resolution input to capture fine details. 
The performance of the GL-FNO is compared with state-of-the-art deep learning methods, including FNO, U-NO, U-FNO, Vision Transformer, CNN-RNN, and CNN-LSTM, to demonstrate its accuracy, computational efficiency, and scalability. Furthermore, physics analysis from domain experts is also performed to demonstrate the reliability of GL-FNO. The results demonstrate that GL-FNO not only accelerates the MHD simulation (a few seconds for prediction, more than ×20,000 speed up) but also provides reliable prediction capabilities, thus greatly contributing to the understanding of space weather dynamics. Our code implementation is available at \url{https://github.com/Yutao-0718/GL-FNO}. 
\end{abstract} 


\section{Introduction}
Studying the solar outer atmosphere has been a longstanding challenge in astrophysics, with its complex magnetic structures playing a crucial role in various solar phenomena \cite{wang2022high,he2023coronal,borovsky2024future}. In the convection zone, the gas pressure dominates the magnetic field pressure, causing the plasma to move and carry the field along with it. These movements generate energetic flows and mass transfer from the chromosphere to the corona. As a result of exertion on the magnetic fields, most of the energy conveyed to the outer solar atmosphere is emitted in the chromosphere. In addition, in the chromosphere, the dynamics transition from being dominated by gas pressure to being dominated by magnetic force. Consequently, understanding and precisely simulating these intricate interactions are important for the progression of our understanding of space weather and its impact on the planet Earth.

Coronal magnetic field modeling is mainly divided into magnetohydrodynamics (MHD) models \cite{braginskii1965transport,pontin2022magnetic,yamasaki2022data,inoue2023comparative}, magnetohydrostatics (MHS) models \cite{shafranov1958magnetohydrodynamical, 2018ApJ...866..130Z}, force-free models \cite{woltjer1958theorem,jing2010nonlinear,jing2014evolution}, and potential-field models \cite{wang1992potential}. These models solve specific partial differential equations (PDEs), ranging from complex magnetohydrodynamics processes to simplified assumptions of current-free conditions, to simulate and understand the structure and dynamics of the coronal magnetic field. These equations are discretized and solved on a grid, but the wide range of spatial and temporal scales and the complex nonlinear interactions between these scales require fine grids and high resolution, making these modeling extremely expensive. 

In recent years, there have been many attempts by researchers to simulate the solar atmosphere by MHD. Leenaarts et al. \cite{leenaarts2007non} simulated the solar atmosphere from the convection zone to the corona. Den \cite{Den:2016jG} presented a three-dimensional MHD simulation code designed to simulate space plasma phenomena with applications in modeling the solar surface and global solar wind structure. Carlsson et al. \cite{carlsson2016publicly} used the Bifrost MHD to provide the community with a realistic simulated magnetic field of the Sun's outer atmosphere by elucidating the complex interactions between the magnetic field and the plasma. The Bifrost MHD is also used for performance evaluation of coronal field reconstruction using various approaches and tools \cite{2017ApJ...839...30F, 2019ApJ...870..101F}. The Bifrost-based MHD model incorporates many physical processes and offers higher spatial and temporal resolution, focusing on the chromosphere and corona. This makes it well-suited for our study of predicting magnetic field topologies in corona.

Coronal magnetic field modeling is extremely time-consuming. To reduce the computational cost, researchers have attempted to simulate the solar outer atmosphere using artificial intelligence (AI)/machine learning (ML) methods. AI/ML has been proven to be efficient in various applications \cite{berg2018unified, han2018solving, long2018pde, malek2006numerical,dan2024evaluation,dan2024multiple, zhang2021developmental}.  Researchers have used AI/ML methods for coronal magnetic field modeling, especially using Physics-Informed Neural Network (PINN)~\cite{raissi2019physics,karniadakis2021physics,gnanasambandam2023self}. PINN incorporates physical laws, typically in the form of PDEs, directly into the neural network's loss function. The network is trained to minimize the residuals of the PDE and satisfy any boundary and initial conditions. For instance, Jarolim et al. \cite{jarolim2023probing} presented an approach for coronal magnetic-field extrapolation, using PINN that integrates observational data and the physical force-free magnetic-field model. The method flexibly finds a trade-off between the observation and force-free magnetic-field assumption, improving the understanding of the connection between the observation and the underlying physics.  To further relax the force-free assumption for the dense plasma conditions in the lower atmosphere, Jarolim et al. \cite{jarolim2024advancing} utilized multi-height magnetic field measurements in combination with PINN to advance solar magnetic field extrapolations. In addition, Baty and Vigon\cite{baty2024modelling} explored the functionality of PINNs through application to two particular solar problems. First, they consider the computation of two-dimensional (2D) force-free magnetic equilibria representative of arcades and loop-like structures in the solar corona by solving an associated Grad–Shafranov-like equation. Second, their method is extended to a more complex system of differential equations that is an incompressible resistive MHD set, to compute 2D magnetic reconnection solutions.  

Despite a significant advancement over traditional physics-based modeling, PINNs have the following issues (1) \texttt{Limited Generalization:} PINNs are designed to solve a specific PDE or a specific instance of a problem. They are well-suited for problems where the PDE is fixed including boundary/initial conditions. For a different PDE, a different PINN needs to be trained for the solution. (2) \texttt{Training Scalability:} Compared to standard neural networks, PINNs often require more complex and time-consuming training processes because they need to satisfy both data-driven and physics-based loss functions \cite{doumeche2023convergence}. 

To address the above issues of PINNs, Fourier neural operator (FNO) \cite{Li2020FourierNO,kovachki2023neural} has been introduced to learn a mapping between function spaces, effectively learning the solution operator for a family of PDEs. Instead of solving a single instance of a PDE, they generalize across different input conditions \cite{azizzadenesheli2024neural}. FNO is trained on a dataset of solutions to the PDE across different parameter settings or initial conditions. Once trained, they can quickly predict solutions for a new instance. A key reason for FNO's success is its ability to accurately model long-range dependencies in spatiotemporal data by learning global convolutions in a computationally efficient manner \cite{bonev2023spherical}. Up to now, there is only one paper using FNO to learn MHD solar wind plasma parameters as output using the solar photospheric magnetic field observation as input \cite{zhao2024prediction}. There is no documented application of FNOs to directly derive the solution of coronal magnetic field modeling (i.e., the Bifrost-based MHD model). 

In this paper, we propose a global-local Fourier Neural Operator (GL-FNO) to accelerate coronal magnetic field modeling (i.e., the Bifrost-based MHD model). GL-FNO contains two branches of FNOs: global FNO takes downsampled input to reconstruct global features while the local FNO takes original resolution input to capture fine details. In our global and local FNOs implementation, we incorporate tucker decomposition \cite{kolda2009tensor, shen2022smooth} to model the kernel operator with reduced memory requirement and better generalization. To summarize, the contributions of this paper are
\begin{itemize}
    \item We proposed a novel method GL-FNO to obtain solutions for coronal magnetic field modeling (i.e., the Bifrost-based MHD model), enabling global and local features learning simultaneously. 
    \item We use data from \cite{carlsson2016publicly} to demonstrate the effectiveness and efficiency of GL-FNO by comparing it with state-of-the-art deep learning methods such as FNO \cite{Li2020FourierNO}, U-NO \cite{rahman2022u}, U-FNO \cite{wen2022u}, Vision Transformer \cite{Dosovitskiy2020AnII}, CNN-RNN \cite{wang2016cnn}, and CNN-LSTM \cite{hara2018can}. Specifically, our GL-FNO achieves the best test accuracy compared to other deep learning methods.
    \item In addition to the evaluation metrics from an AI/ML perspective, evaluation from a physics perspective is conducted to verify that our prediction from GL-FNO is reliable.
\end{itemize}

 The rest of this paper is organized as follows.  The Bifrost-based MHD Model is introduced in Section~\ref{sec: problem setting}, followed by our proposed GL-FNO in Section~\ref{sec: methods}. Our experiments and results are summarized in Section~\ref{sec: experiments}. Finally, conclusions are discussed in Section~\ref{sec: conclusion}.

\begin{figure*}[!htbp]
	\centering
	\includegraphics[width=1\linewidth]{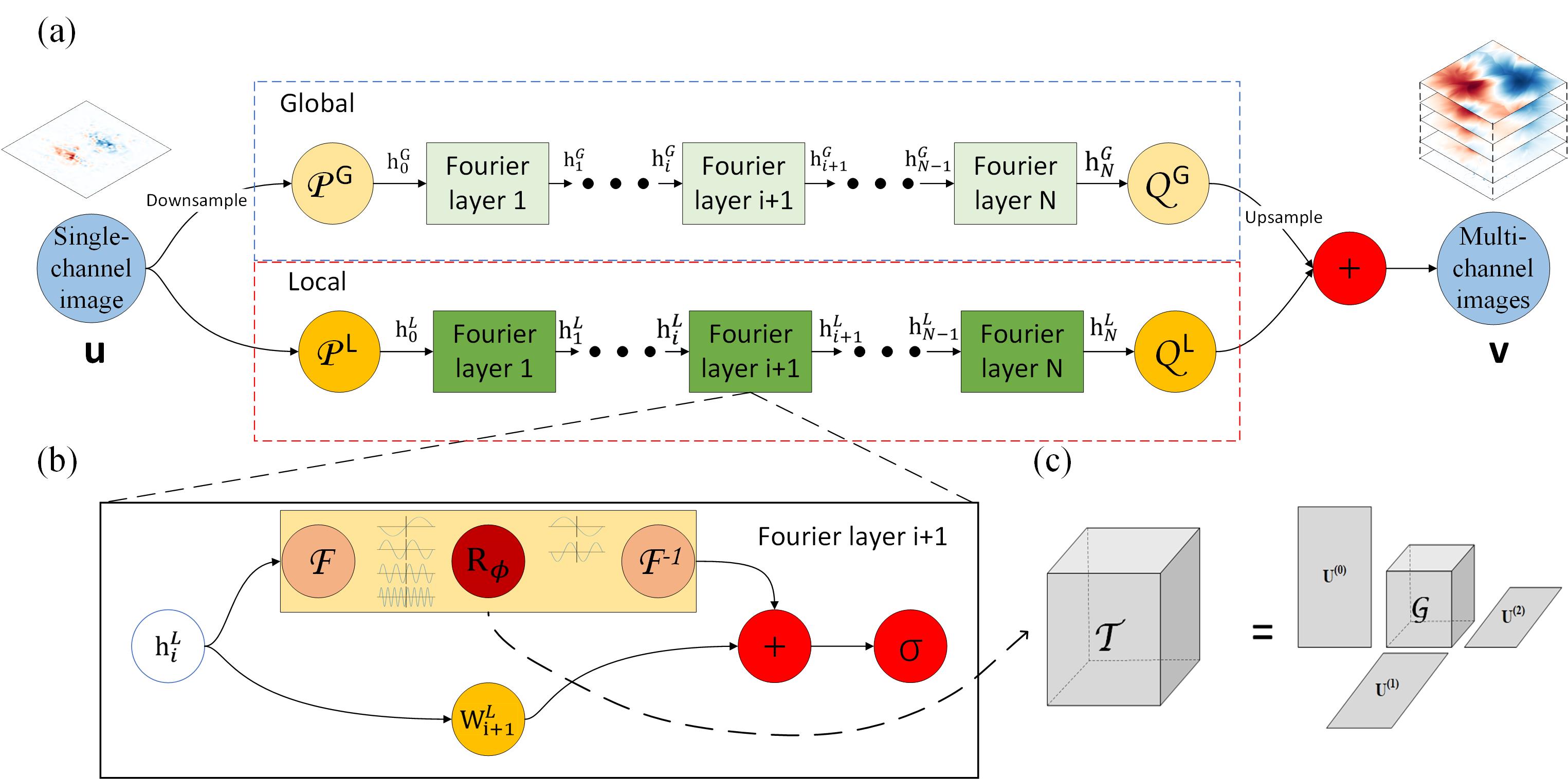}	
	\caption{\textbf{(a)} The architecture of the global-local fourier neural operators; \textbf{(b)} local fourier layer; \textbf{(c)} tucker decomposition. }
	\label{GL-FNO}       
\end{figure*}
\section{Bifrost-based MHD Model} \label{sec: problem setting}
Bifrost is a flexible and massively parallel code described in detail in \cite{gudiksen2011stellar}. The Bifrost-based MHD model represents the realistic simulations of the magnetic solar outer atmosphere, with the 3D magnetic topologies on an enhanced area on the Sun. Bifrost has evolved from earlier numerical codes developed by \cite{Nordlund1995VisualizingA3} and \cite{galsgaard1996heating}, these codes share a common core. One of the key features of the Bifrost-based MHD model is its ability to simulate the multi-scale dynamics of solar phenomena, ranging from small-scale magnetic reconnection events to large-scale coronal mass ejections (CMEs). It incorporates realistic physics processes, such as radiative transfer, non-equilibrium ionization, and thermal conduction, to provide a comprehensive understanding of solar atmospheric phenomena. Specifically, a staggered mesh explicit code that solves the standard MHD PDEs on a Cartesian grid as follows 
\begin{align}
\frac{\partial \rho}{\partial t} &= -\nabla \cdot (\rho \mathbf{u}) \\
\frac{\partial (\rho \mathbf{u})}{\partial t} &= -\nabla \cdot (\rho \mathbf{u} \mathbf{u} - \tau) - \nabla P + \mathbf{J} \times \mathbf{B} + \rho \mathbf{g} \\
\mu \mathbf{J} &= \nabla \times \mathbf{B} \\
\mathbf{E} &= \eta \mathbf{J} - \mathbf{u} \times \mathbf{B} \label{E}\\
\frac{\partial \mathbf{B}}{\partial t} &= -\nabla \times \mathbf{E} \label{B}\\
\frac{\partial e}{\partial t} &= -\nabla \cdot (e\mathbf{u}) - P\nabla \cdot \mathbf{u} + Q
\end{align}
where $\rho$, $\mathbf{u}$, $e$, $\mathbf{B}$ are the density, the velocity vector, the internal energy per unit volume, and the magnetic field intensity respectively. $\tau$, $P$, $\mathbf{J}$, $\mathbf{g}$, $\mu$, $\mathbf{E}$ and $\eta$ are the stress tensor, the gas pressure, the electric current density vector, the gravitational acceleration, the vacuum permeability, the electric field vector and the magnetic diffusivity respectively. The quantity $Q$ can contain many terms, depending on the individual experiment. It could for instance contain a term from the chosen Equation Of State (EOS), a term containing the effect of the Spitzer thermal conductivity, a term from radiative transfer, etc. The EOS needed to close this set of equations can be anything from a simple ideal gas EOS to a complex EOS including detailed microphysics \cite{gudiksen2011stellar}. 

The process of modeling the outer solar atmosphere is explained in \cite{gudiksen2011stellar}. Consider the example of the MHD PDE equation for the magnetic field intensity $\mathbf{B}$. Combining \eqref{E} and \eqref{B} results in the MHD Magnetic Induction Equation:
\begin{align}
    \frac{\partial \mathbf{B}}{\partial t} &= \nabla \times (\mathbf{u} \times \mathbf{B}) - \nabla \times (\eta \mathbf{J}).
\end{align}
To solve this equation, Bifrost employs a staggered grid discretization approach using the finite difference method \cite{nordlund19953d}, which assigns the different components of the magnetic field and velocity to different locations within the grid cells. Subsequently, the curl of the magnetic field is computed using a sixth-order exact scheme to handle spatial derivatives. The velocity field and magnetic field interaction term $\mathbf{u} \times \mathbf{B}$ is interpolated to the desired grid location, after which the curvature is calculated. Bifrost employs either a third-order Runge-Kutta method or a third-order Hyman method for the calculations to advance the solution of the differential equations in time. Finally, Bifrost implements non-periodic boundary conditions using “ghost zones” \cite{gudiksen2011stellar}. Depending on the specific problem, different types of boundary conditions (e.g. symmetric, asymmetric, or extrapolated boundary values) are applied to these ghost zones. After performing these calculations, Bifrost can obtain the solution for the magnetic field $\mathbf{B}$.

\section{Proposed Method} \label{sec: methods}
Our proposed GL-FNO is a modification of FNO. In GL-FNO, we have a global iterative architecture: $h^G_0 \mapsto h^G_1 \mapsto \dots \mapsto h^G_N$ and a local iterative architecture: $h^L_0 \mapsto h^L_1 \mapsto \dots \mapsto h^L_N$, where $\{h^G_i\}_{i=0}^N,\{h^L_i\}_{i=0}^N$ are a sequence of global and local functions, respectively. As shown in Fig.~\ref{GL-FNO}(a), the input $u(\textbf{x})$ is first lifted to a higher dimensional global representation $h^G_0=\mathcal{P}^G(u(\textbf{x}))$ and local representation $h^L_0=\mathcal{P}^L(u(\textbf{x}))$ by the transformations $\mathcal{P}^G$ and $\mathcal{P}^L$, which are usually parameterized by shallow fully-connected neural networks. Then we apply several iterations of global update $h^G_i\mapsto h^G_{i+1}$ and local update $h^L_i\mapsto h^L_{i+1}$ (introduced later). The output $v(\textbf{x})=\mathcal{Q}^G(h^G_N(\textbf{x}))+\mathcal{Q}^L(h^L_N(\textbf{x}))$ is the summation of the projections by the transformations $\mathcal{Q}^G$ and $\mathcal{Q}^L$.

Each iterative update is enabled by Fourier layers as 
\begin{equation}
    \begin{aligned}
        \texttt{Global:} h_{i+1}^G&=\sigma\left(\mathcal{W}^G_{i+1}h_i^G(\textbf{x})+\mathcal{K}^G_{\theta^G_{i+1}}h_i^G(\textbf{x})\right),\\
        \texttt{Local:} h_{i+1}^L&=\sigma\left(\mathcal{W}^L_{i+1}h_i^L(\textbf{x})+\mathcal{K}^L_{\theta^L_{i+1}}h_i^L(\textbf{x})\right),
    \end{aligned}
\end{equation}
where $\sigma$ is a nonlinear activation function, $\mathcal{K}^{G}_{\theta^G_{i+1}}$ or $\mathcal{K}^{L}_{\theta^L_{i+1}}$ is the kernel integral operator parameterized by a neural architecture with learnable parameters $\theta^G_{i+1}$ or $\theta^L_{i+1}$. 

For a better understanding of the structure of a Fourier layer, we visualize the Fourier layer of the local branch in Fig.~\ref{GL-FNO}(b).  
The input at the Fourier layer $i+1$, i.e., $h_i^L(\textbf{x})$, undergoes two transformations. \textbf{In the top path}, $h_i^L(\textbf{x})$ is transformed to the Fourier space through Fourier transform $\mathcal{F}$ followed by a linearly transform $\mathcal{R}$ filters out the high-level Fourier modes, and finally the inverse Fourier transform $\mathcal{F}^{-1}$ converts back into the original space. Unlike \cite{li2020neural} in which message passing is used as in a graph neural network, the Fourier layer leverages the convolution theorem, which states convolution is equivalent to point-wise multiplication in the Fourier space. Therefore, the Fourier layer initializes complex-valued learnable parameters directly in the Fourier space. Following the notations in \cite{Li2020FourierNO}, Fourier integral operator is defined by
\begin{equation}
    \mathcal{K}^L_{\theta}h^L_i = \mathcal{F}^{-1}\left(\mathcal{F}(\mathcal{K}^L_{\theta}\right) \cdot \mathcal{F}(h_i^L)) = \mathcal{F}^{-1}\left(R_{\phi}\mathcal{F}(h^L_i)\right)
\end{equation}
where $R_{\phi}$ is the Fourier transform of  $\mathcal{K}^L_{\theta}$, i.e., $R_{\phi}=\mathcal{F}(\mathcal{K}^L_{\theta})$, and output is real-valued in the latent space. In the implementation, the fast Fourier transform and its inverse is used to preserve computational efficiency. \textbf{The bottom path} applies $h^L_i(x)$ with a linear transform $\mathcal{W}^L_{i+1}$, where the discretized $\mathcal{W}^L_{i+1}$ is a matrix of learnable parameters. The results of the two paths are then added together, followed by a nonlinear activation function $\sigma$.

In GL-FNO, we also parameterize the learnable weights within the Fourier domain with a low-rank tucker factorization as shown in Fig.~\ref{GL-FNO}(c). This factorization leads to a significant reduction in the number of parameters required to represent the operator, enhancing the model's data efficiency and generalization capabilities \cite{Kossaifi2023MultiGridTF}.

Our GL-FNO learns the operator to map from input space $u$ to the solution space $v$ with global and local branches. The global branch captures the overall structure and dynamics, while the local branch focuses on fine-grained details. This fusion enables the model to generalize better across different physical conditions while retaining accuracy in areas requiring detailed resolution. Additionally, by integrating both scales, the network can learn complex interactions more efficiently, leading to faster and more accurate prediction in diverse physical systems.

\section{Experiments and Results} \label{sec: experiments}
All data used in the experiments are available from Hinode Science Data Centre Europe (\url{https://sdc.uio.no/search/simulations}). In our study, we will focus on the 3D magnetic data cube  $\textbf{B}$ along the $x$, $y$, and $z$ axes—denoted as $B_x$, $B_y$, and $B_z$, respectively. These components depict the 3D magnetic environment of the region of interest, in our case, an enhanced network area on the Sun. The data are located in the ``en024048\_hion", a collection of 3D simulated magnetic data cubes generated using the Bifrost code to simulate a computational volume with a magnetic field topology similar to an enhanced network area on the Sun \cite{2016A&A...585A...4C}. We use the dataset of $B_z$ as one example to illustrate the data formats (the same applies to $B_x$ and $B_y$). The first snapshot of $B_z$ was captured at $t=3850$ seconds and the last snapshot of $B_z$ was captured at $t=5410$ seconds. Snapshots were recorded at 10-second intervals, resulting in 157 cubes. Therefore, $B_z$ dataset contains $\{B_z^j\}_{j=1}^{157}$.  The size of each cube $B_z^j$ is $504 \times 504 \times 496$.  The third dimension (i.e., 496) represents the height, which extends from the upper convection zone to the corona and is aimed at aiding the study of the solar chromosphere, a region that is challenging to model due to its complex dynamics and physics \cite{carlsson2016publicly}. The simulation volume measures 24 Mm $\times$ 24 Mm horizontally ($504 \times 504$ pixels), with a depth of 2.4 Mm below the solar surface, and extends 14.4 Mm above, covering the upper part of the convection zone, the photosphere, chromosphere, transition region, and corona. 

\subsection{Data Preprocessing} 
Given the constraints (limited memory space) of our GPU server, we downsampled the cube from $504 \times 504 \times 496$ to $504 \times 504 \times 100$ by selecting every five slices with index $1,6,11,\dots,496$. \textbf{Our goal here is to learn the mapping from the bottom-most slice (the boundary condition of PDEs) to the remaining 99 slices as shown in Fig.~\ref{GL-FNO} (a)}. The deep learning method inputs the bottom-most image of size $504 \times 504$ and outputs a 3D cube of size $504 \times 504 \times 99$. In addition, data normalization is implemented. 90\% of the data is randomly selected for training. 

\begin{figure*}[!htbp]
	\centering
	\includegraphics[width=0.75\linewidth]{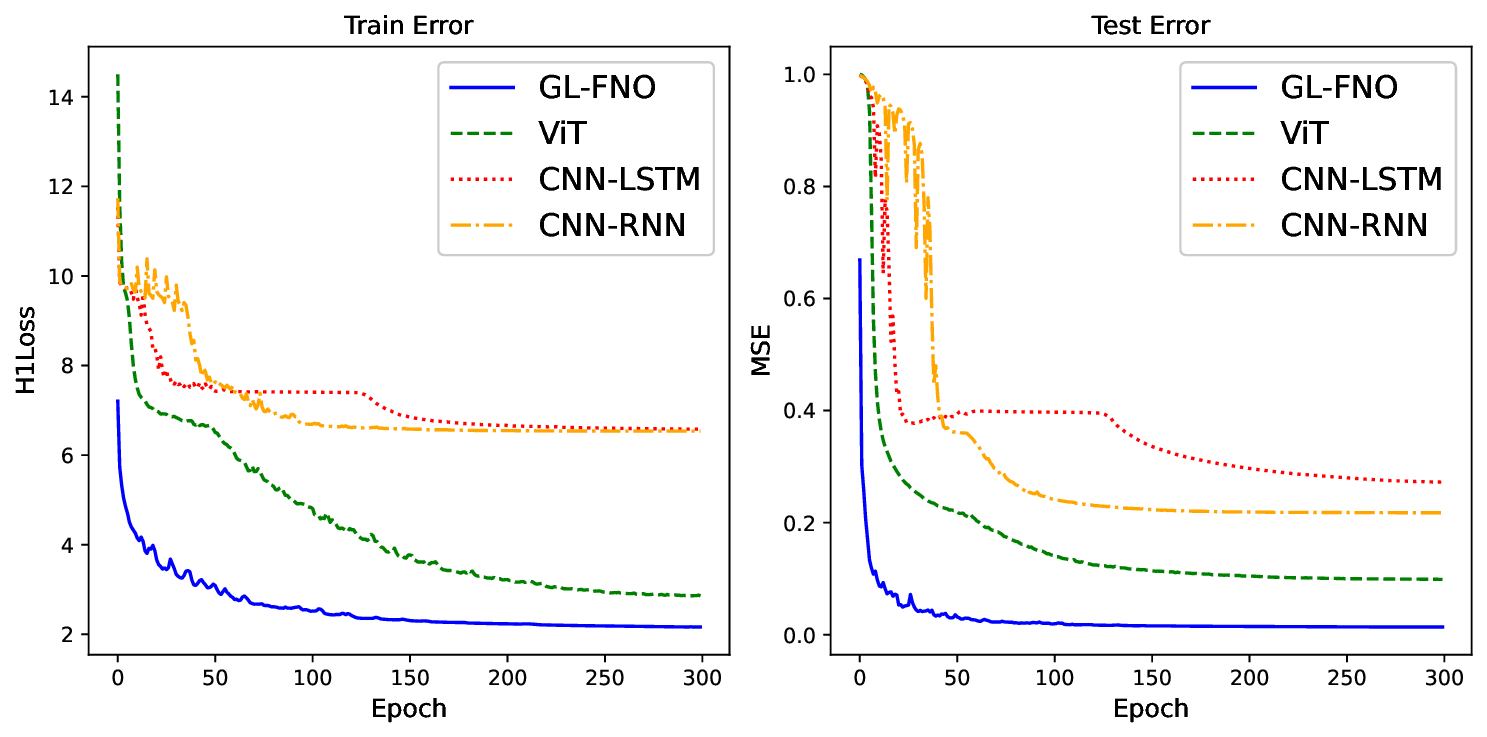}	
	\caption{Training and test error curves for different epochs of GL-FNO, ViT, CNN-RNN, and CNN-LSTM.} 
	\label{linegraphy}       
\end{figure*}
\subsection{Experiment Settings}
In our experiments, we implemented three other deep learning methods together with GL-FNO. 
\subsubsection{GL-FNO Structure} \label{subsubsec: GL-NFO}
Our GL-FNO is built based on publicly available code (\url{https://github.com/neuraloperator/neuraloperator}). For the local branch, we set the number of modes retained in each dimension in the Fourier layer to $(64,64)$. We set the hidden channel to $128$, thus enabling the model to efficiently address the complex spatial structure inherent in the data. The model uses the Tucker decomposition with a rank$=0.5$, which balances expressiveness with computational and storage efficiency. It captures the most important interactions between input data elements while discarding less important ones to create a more compact model. In addition to the basic Lifting Layer, FNO Blocks, and Projection Layer mentioned in Section~\ref{sec: methods},  a layer of Multi-Layer Perceptrons (MLPs) is added after the FNO Blocks to output the desired size of the 3D cube ($504 \times 504 \times 99$) since the original code can only output 2D image. Additionally, it enables an additional non-linear processing stage, improving its capability to model complex functions and enhancing its ability to learn a diverse array of PDE solutions. We keep the same structure for the global branch as the local one.

\subsubsection{Vision Transformer (ViT)}
The ViT \cite{Dosovitskiy2020AnII} used in our paper is a modification of the transformer architecture traditionally used for Natural Language Processing (NLP). Our ViT is adapted to process the input image in the form of patches, which are embedded with masks and then passed through the transformer encoder. The model processes $504\times504$ images using $8\times8$ patches. With 128-dimensional embedding and a multi-head attention mechanism spanning 2 layers of 8 heads. A learnable mask embedding is added to the patch embedding sequence, and positional embedding is added to the spatial information lost during the patch embedding process. After converter processing, the output is summarized and decoded to the desired number of output channels for our prediction.

\subsubsection{CNN-RNN and CNN-LSTM}
We use two different encoder-decoder structures: The CNN-RNN \cite{wang2016cnn} and CNN-LSTM \cite{hara2018can}. We use Convolutional Neural Networks (CNN) for feature extraction and Recurrent Neural Networks (RNN) or Long Short-Term Memory Networks (LSTM) for modeling sequential images. The CNN component consists of three convolutional layers that progressively increase the number of feature channels from 1 to 256 while down-sampling the image dimensions by a factor of 2 after each layer. This is done via convolution with a stride of 2, effectively reducing the spatial dimensions and extracting higher-level features. In CNN-RNN, the main component of the recurrent network is a series of gated recurrent units (GRUs), which are chosen for their ability to efficiently model time series. The hidden state size of the GRU units in this model is 256, which reflects a balance between model complexity and computational efficiency, allowing the network to capture a large amount of temporal information without being too computationally demanding. Meanwhile, the CNN-LSTM uses LSTM cells with a hidden state size of 256, which is consistent with the output dimension of the CNN.

\subsubsection{Optimizer and Scheduler}

All deep learning methods are trained using the Adam optimizer \cite{Kingma2014AdamAM} and the ReduceLROnPlateau scheduler with 300 epochs and a batch size of 10.  An Adam optimizer was used with an initial learning rate of 0.001 and a weight decay of 0.0001. The ReduceLROnPlateau scheduler was used to adjust the learning rate according to the performance of the model, with a learning rate reduction factor of 0.8 and patience of 2. Huber Loss \cite{taylor2023deep} is used for training for all the methods to approximate PDE solutions. It is a hybrid loss function, known for blending the properties of mean squared error and mean absolute error, demonstrating robustness against outliers—a feature when dealing with the intrinsic discontinuities and sharp gradients characteristic of PDEs.

\subsection{Results}
We mainly aim to address two questions
\begin{itemize}
    \item Section~\ref{subsubsec: AI/ML}: What is the performance of GL-FNO compared with other deep learning methods?
    \item Section~\ref{subsubsec: physics}: Do the solutions (i.e., prediction) generated from GL-FNO make sense from the perspective of a solar physicist?
\end{itemize}
\subsubsection{Evaluation from an AI/ML Perspective}  \label{subsubsec: AI/ML}
Fig.~\ref{linegraphy} shows the training error, i.e., Huber Loss, and test error, i.e., Mean Squared Error (MSE), over 300 epochs for the GL-FNO, ViT, CNN-RNN, and CNN-LSTM. The training and test errors of GL-FNO and ViT show an overall decreasing trend and converge to a much lower error than CNN-RNN and CNN-LSTM. GL-FNO not only converges to the lowest training and test error but also shows very stable and smooth training and test error curves compared to all other deep learning methods.  

\begin{table}[!htbp]
    \caption{Test performance comparison of GL-FNO, ViT, CNN-RNN, and CNN-LSTM for $\textbf{B}$ in terms of MSE, $R^2$, RE, MAE, PSNR, SSIM.}
    \label{table1}       
    \centering
    {\scriptsize
    \begin{tabular}{ccccccc}
        \toprule
        Model &  MSE $\downarrow$& $R^2 	\uparrow$ & RE $\downarrow$& MAE $\downarrow$ & PSNR $\uparrow$ & SSIM $\uparrow$\\ 
        \midrule
        \textbf{GL-FNO} &  \textbf{0.0331} & \textbf{0.9669} & \textbf{0.1756} & \textbf{0.0672}  & \textbf{50.58} & \textbf{0.9883} \\ 
        ViT &  0.1240 & 0.8759 & 0.3498 & 0.1417 &35.42 & 0.8856\\ 
        CNN-RNN &  0.2514 & 0.7485 & 0.4995 & 0.2230 & 30.72 & 0.7781\\ 
        CNN-LSTM   & 0.3257 & 0.6743 & 0.5686 & 0.2940 & 28.91 & 0.7356\\ 
        \bottomrule
    \end{tabular}
    }
\end{table}


\begin{figure}[!htbp]
	\centering
 	\subfloat[$B_x$ Visualization]{\includegraphics[width=0.91\linewidth]{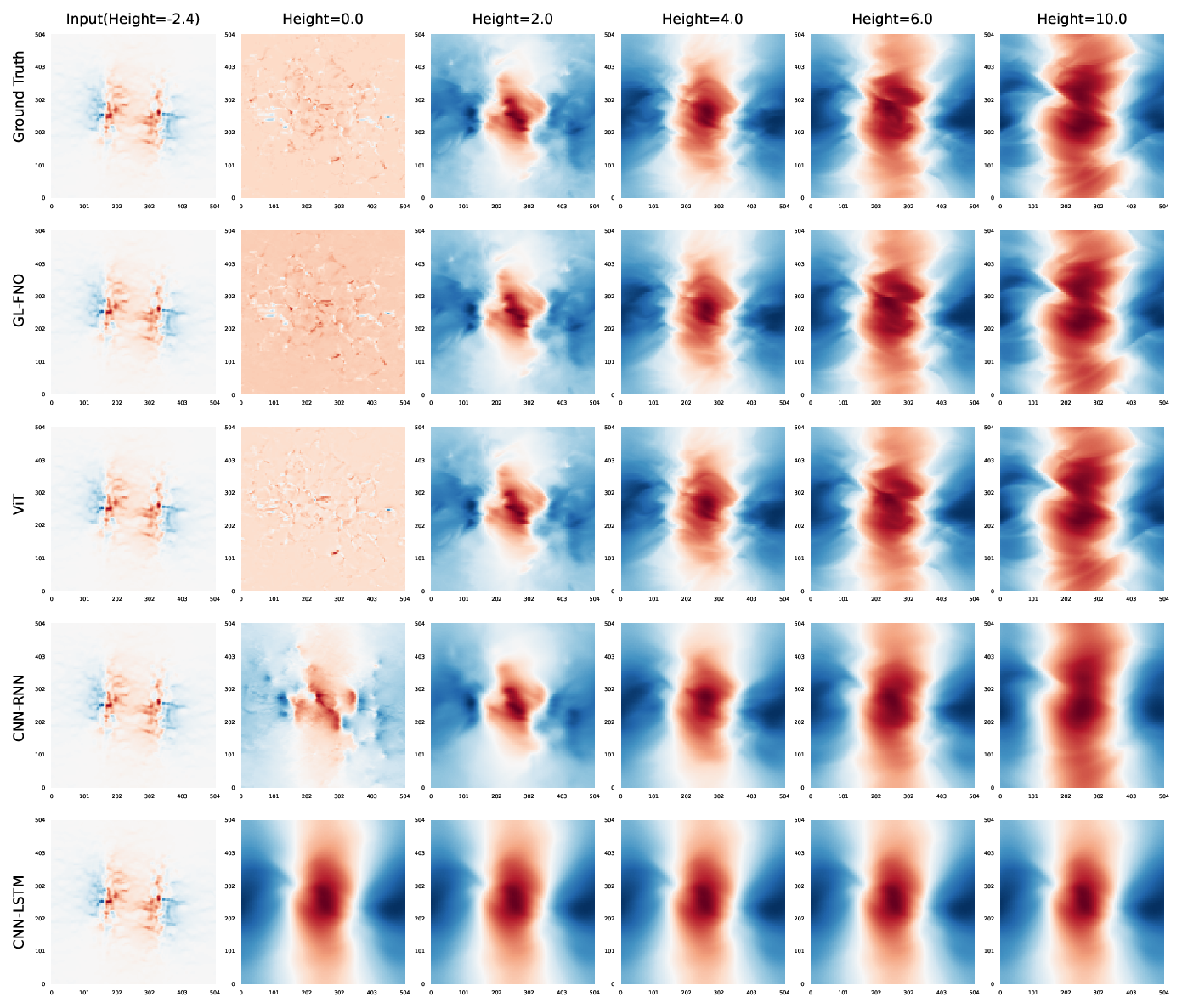} \label{Bx visualization}  }\\
     \subfloat[$B_y$ Visualization]{\includegraphics[width=0.91\linewidth]{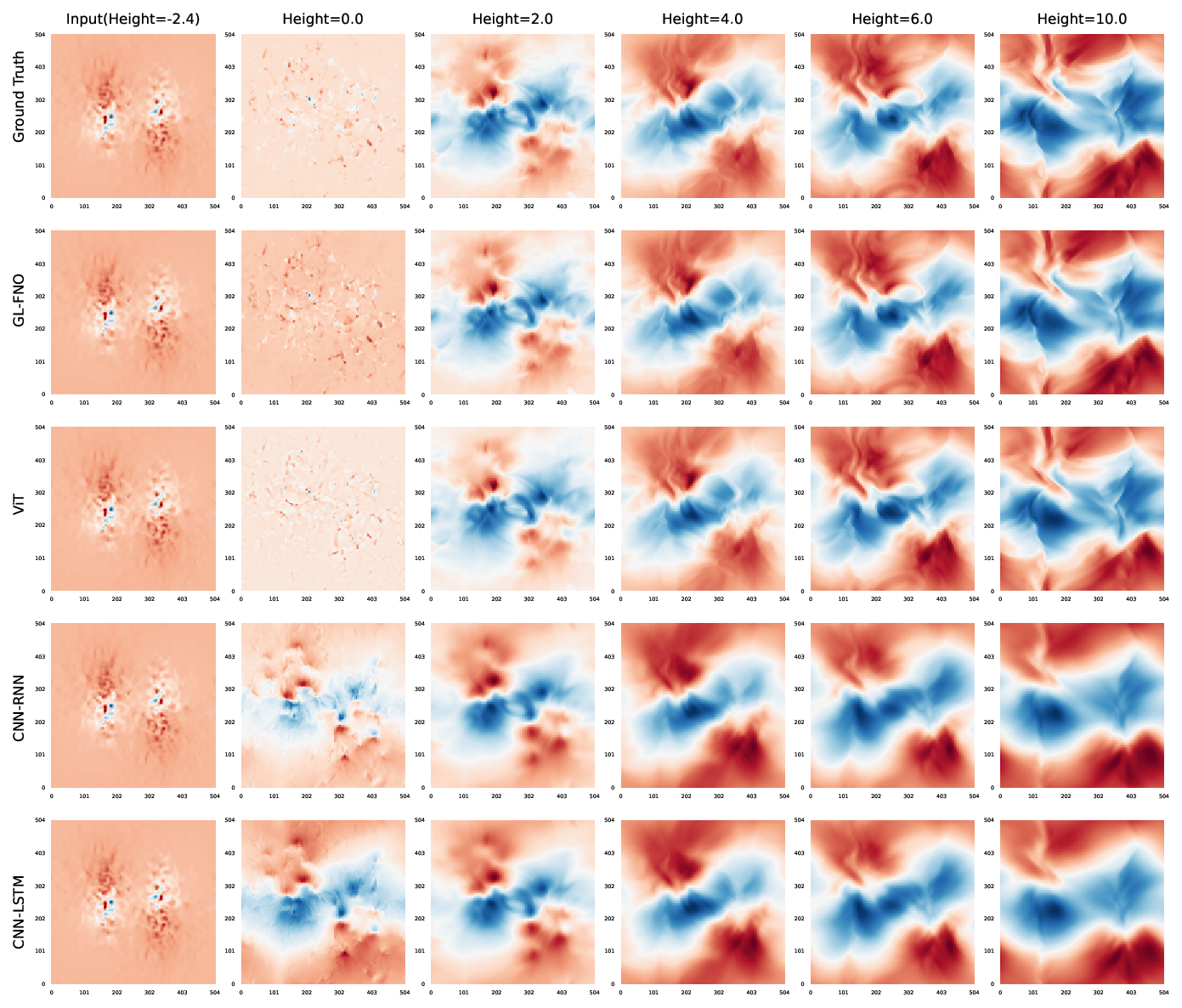} \label{By visualization} }\\
    \subfloat[$B_z$ Visualization]{\includegraphics[width=0.91\linewidth]{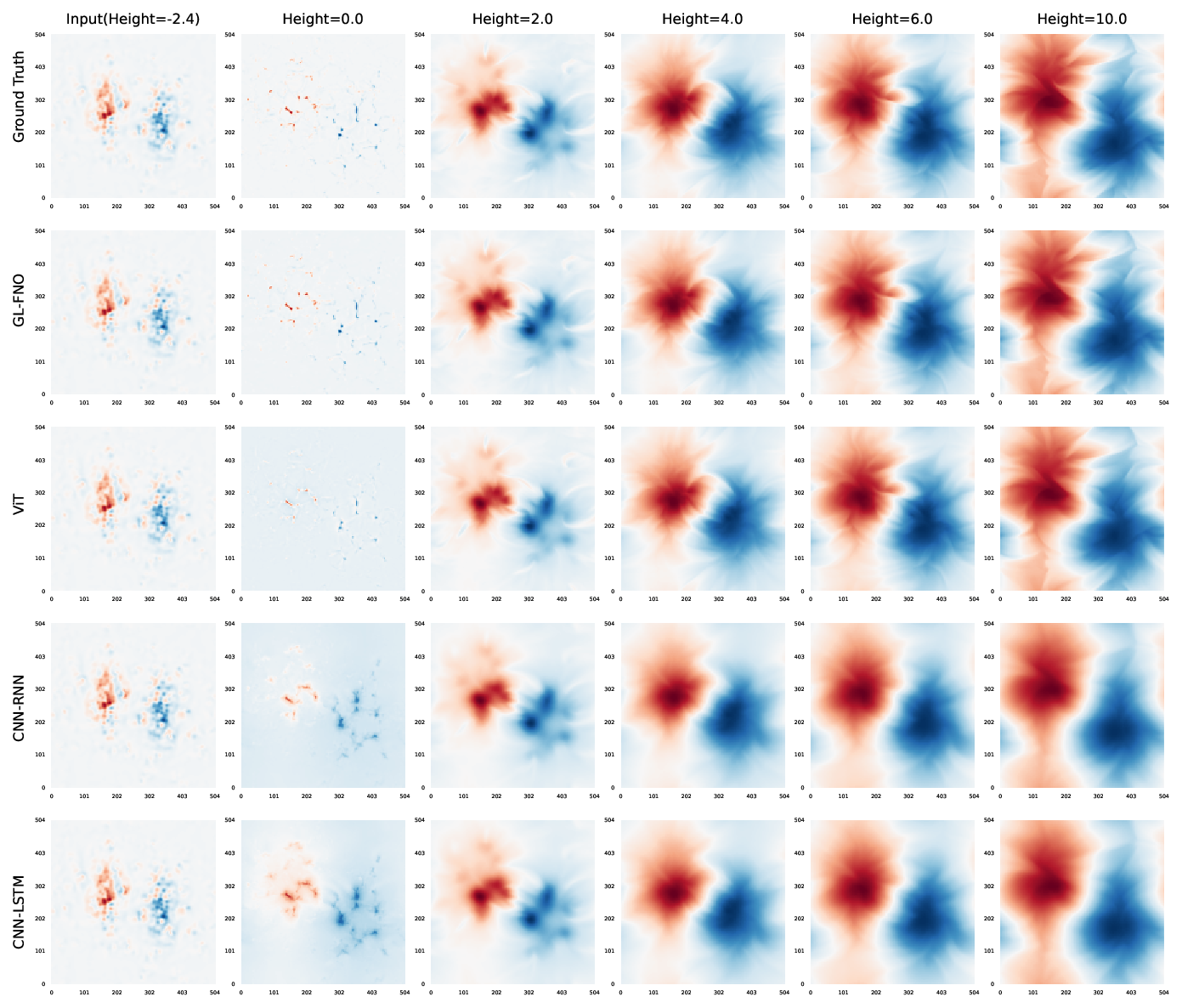} \label{Bz visualization} }
	\caption{The visualization of prediction from GL-FNO, ViT, CNN-RNN, and CNN-LSTM, compared with ground truth at different heights: (a) $B_x$; (b) $B_y$; (c) $B_z$. } \label{fig: vis} 	   
\end{figure}
\begin{figure}[!htbp]
	\centering
  	\subfloat[$B_x$ Error Map]{\includegraphics[width=0.92\linewidth]{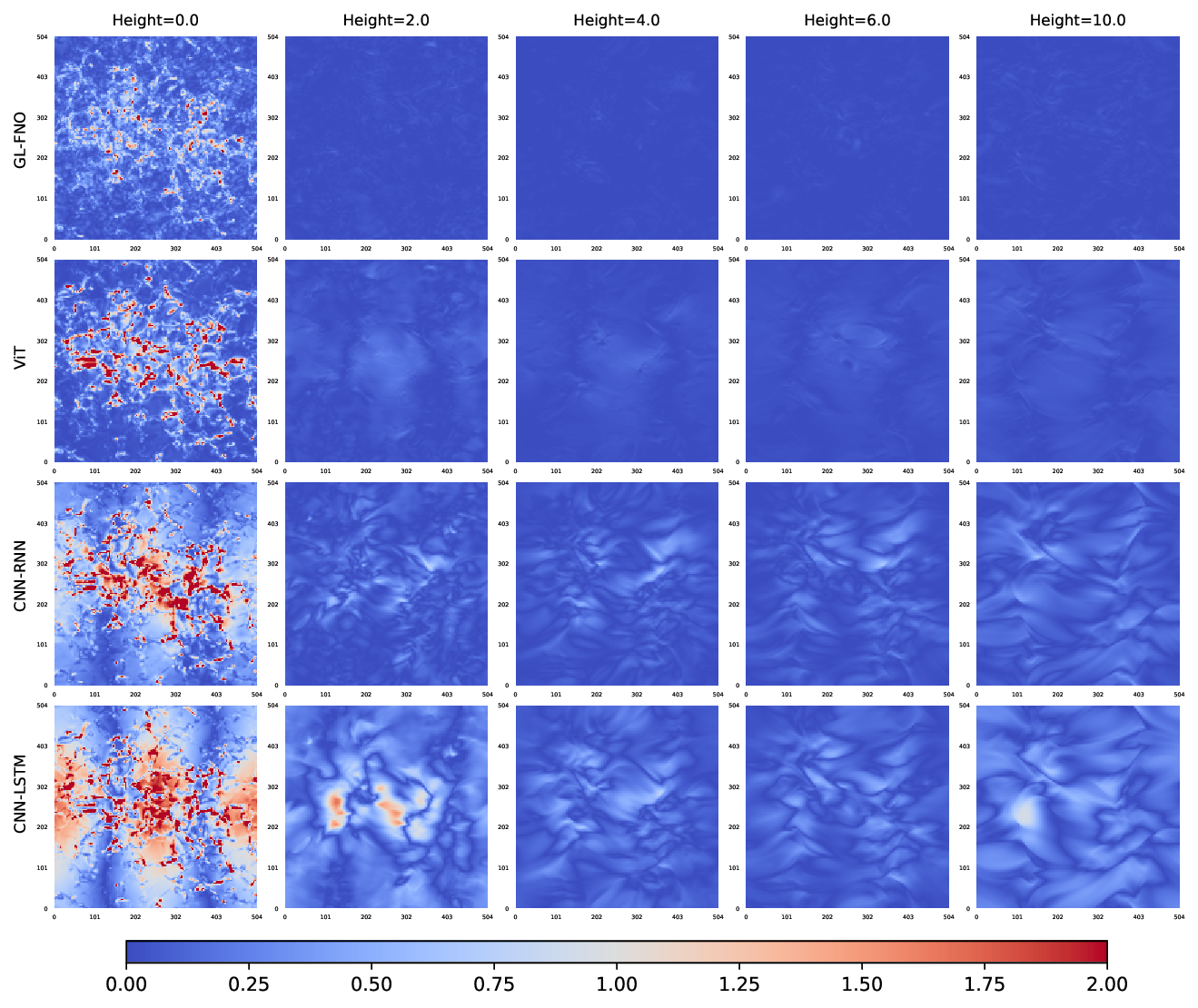} \label{Bx error map}     }\\
     \subfloat[$B_y$ Error Map]{\includegraphics[width=0.92\linewidth]{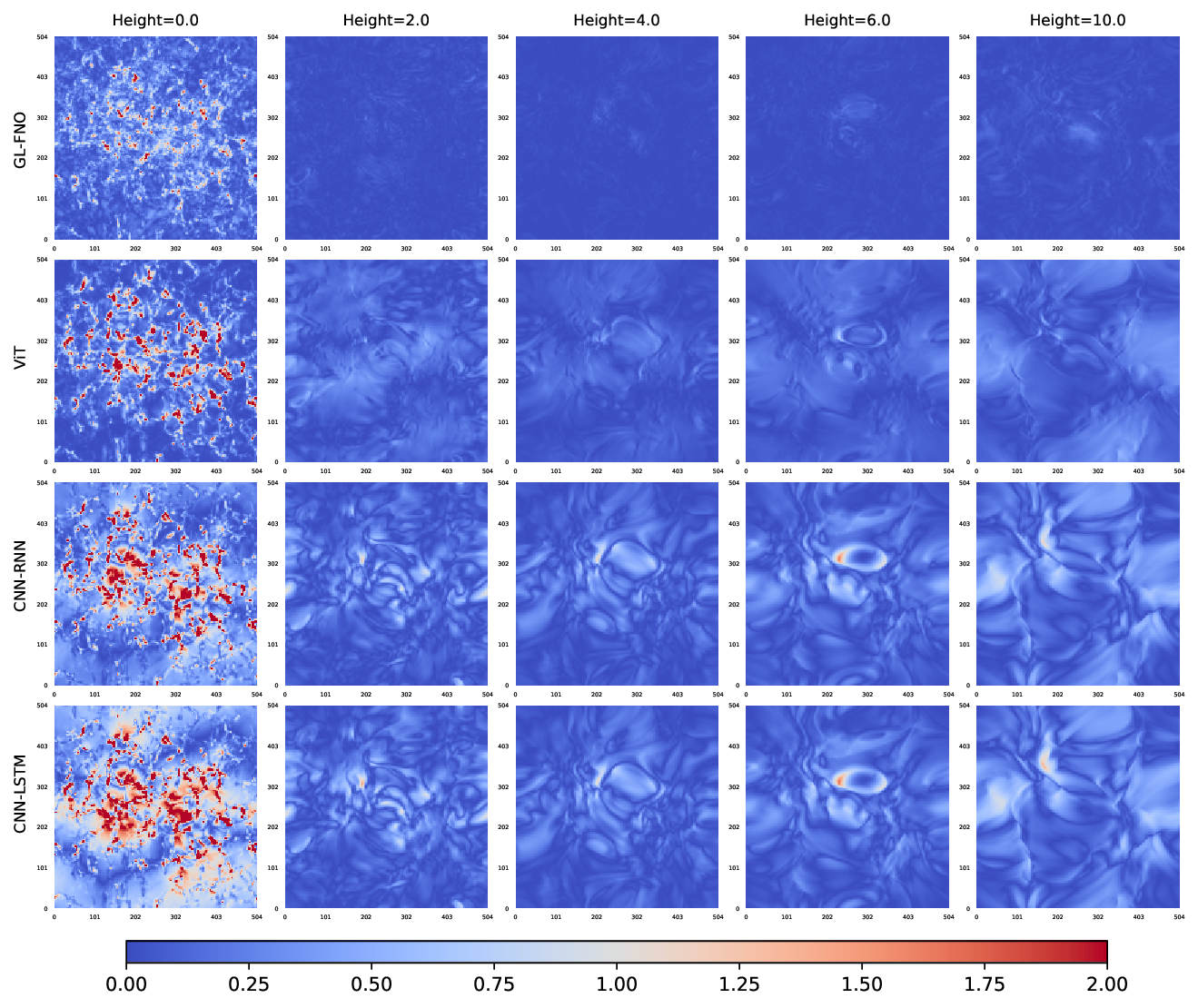} \label{By error map}    }\\
    \subfloat[$B_z$ Error Map]{\includegraphics[width=0.92\linewidth]{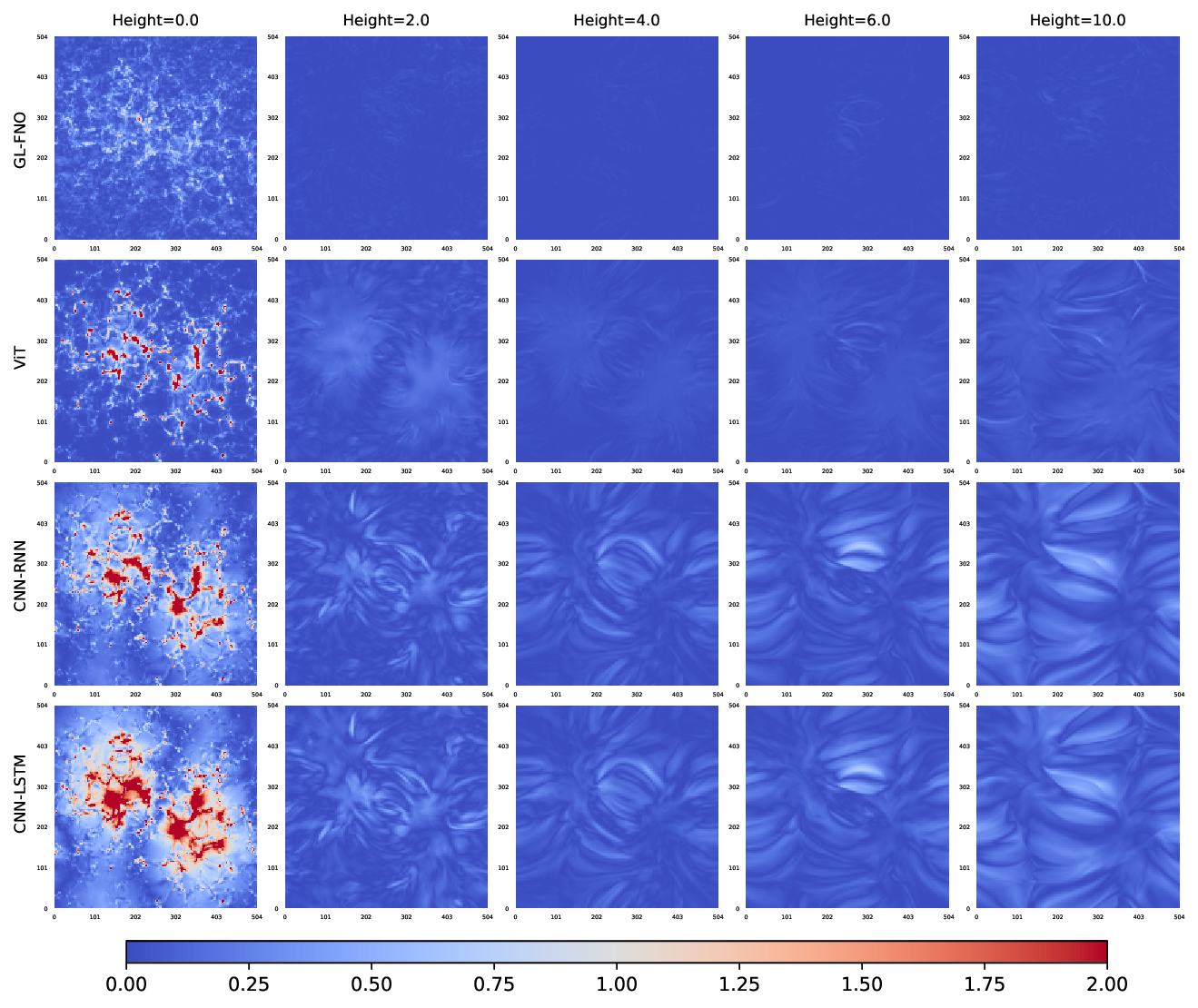} \label{Bz error map}    }
	\caption{The visualization of error maps from GL-FNO, ViT, CNN-RNN, and CNN-LSTM at different heights: (a) $B_x$; (b) $B_y$; (c) $B_z$.  } \label{fig: error maps} 
\end{figure}

In addition to MSE, Table \ref{table1} summarizes the performance of different deep learning methods in terms of R-square ($R^2$), Relative Error (RE), Mean Absolute Error (MAE), Peak Signal-to-Noise Ratio (PSNR), and Structural Similarity Index Measure (SSIM). GL-FNO shows the best test performance among all deep learning methods. GL-FNO has the lowest MSE, RE, and MAE and the highest $R^2$, PSNR, and SSIM. GL-FNO achieves $R^2=0.9498$, PSNR = 50.58, and SSIM = 0.9883 while the ViT achieves the second-best $R^2=0.8759$, PSNR = 35.42, and SSIM = 0.8856. On the other hand, the CNN-RNN and CNN-LSTM (i.e., CNN-based recurrent architectures) show the second-worst and worst test performance, respectively, which are not comparable with GL-FNO and ViT. Overall, GL-FNO shows the best test performance. In our Appendix~\ref{appendix}, we also compared different neural operators with our GL-FNO, where the same conclusion can be obtained.  

Fig.~\ref{fig: vis} shows the visualization of prediction from GL-FNO, ViT, CNN-RNN, and CNN-LSTM, compared with ground truth at different heights. In Figs \ref{Bx visualization}, \ref{By visualization}, and \ref{Bz visualization}, the selected slices are captured at different heights of the cube. The first column in each Figure is the input data of the deep learning methods. The second through fifth columns correspond to heights of approximately 0 Mm, 2 Mm, 4 Mm, 6 Mm, and 10 Mm, respectively. These slices were chosen to demonstrate the diversity of magnetic field strengths at different heights. The first row represents the ``ground truth'', which serves as a benchmark for evaluating the subsequent rows -- GL-FNO, ViT, CNN-RNN, and CNN-LSTM. From these three figures, we can observe that the prediction results of GL-FNO and ViT are almost the same as the ground truth while CNN-RNN and CNN-LSTM reveal remarkable discrepancies in fine details when compared with the ground truth at heights = 0, 2 Mm, 4 Mm, 6 Mm, and 10 Mm. Specifically,  CNN-RNN and CNN-LSTM cannot learn the pattern of the ground truth at heights = 0 Mm.

To better visualize the predictions of different deep learning methods, Fig.~\ref{fig: error maps} shows the error maps of $B_x$, $B_y$, and $B_z$ from GL-FNO, ViT, CNN-RNN, and CNN-LSTM at different heights of approximately 0 Mm, 2 Mm, 4 Mm, 6 Mm, and 10 Mm. These different heights are corresponding to the first, second, third, fourth, and fifth columns, respectively. These three figures are obtained by subtracting the predicted results from the ground truth and then taking the absolute value. The blue color represents a small error, indicating that the predicted values are closer to the actual values. These regions can be interpreted as areas where the deep learning method has reached a high level of accuracy, successfully capturing the underlying patterns and dynamics of the data. Conversely, the red color represents a large error, indicating an error between the output from the deep learning method and the ground truth. The darkness of the red color is proportional to the magnitude of the error, with darker red indicating a larger error. 

Fig.~\ref{fig: error maps} shows that there are prediction errors for all deep learning methods around a height of approximately 0 Mm. However, GL-FNO and ViT still perform much better compared to CNN-RNN and CNN-LSTM. This observation can be attributed to several potential factors, including the inherent complexity of the data. While it may be challenging to find differences between GL-FNO and ViT in Figs \ref{Bx visualization}, \ref{By visualization}, and \ref{Bz visualization}, Figs \ref{Bx error map}, \ref{By error map}, and \ref{Bz error map} allow us to observe that GL-FNO only has the smallest prediction errors at heights of approximately 2 Mm, 4 Mm, 6 Mm, and 10 Mm, where the error map is mostly dark blue (i.e., no errors).  ViT shows serpentine, thin-line errors, indicating a deficiency in capturing fine details as compared to GL-FNO. The CNN-RNN and CNN-LSTM exhibit learning capabilities but present significantly larger errors than GL-FNO and ViT, which are not only quantitatively substantial but also lack meaningful physical interpretation.

In conclusion, GL-FNO stands out for its robustness and ability to capture detailed features with the lowest test error, indicating its potential suitability for complex tasks requiring high precision. In contrast, while ViT shows promise in certain aspects, its performance is shadowed by its shortcomings in complex structure predictions. 

\begin{figure}[!htbp]
    \includegraphics[width=1\linewidth]{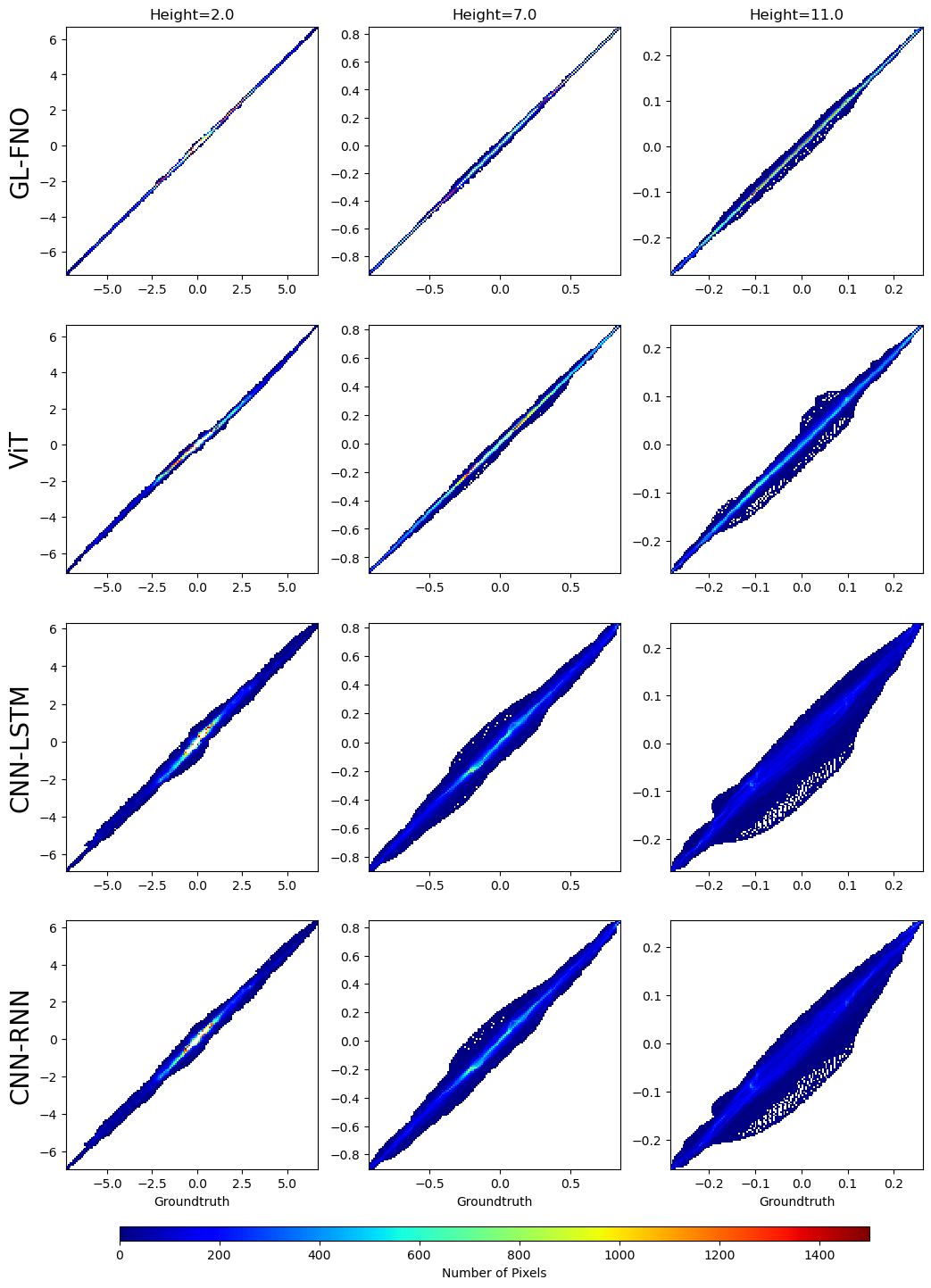}
\caption{2D histogram of GL-FNO, ViT, CNN-RNN, and CNN-LSTM to the ground truth, at height = 2.0 Mm, 7.0 Mm, and 11.0 Mm. Color gradient denotes the number of data points.}  
	\label{2dhist}       
\end{figure}

\subsubsection{Evaluation from a Physics Perspective} \label{subsubsec: physics}

When assessing the performance of deep learning methods in reproducing the MHD outputs, it is vital to check whether these methods can successfully reproduce the comprehensive physical quantities found in the AI-generated data, similar to the ground truth. We examine key physical quantities--specifically, the magnetic field strength and current density as a function of height, to see the capabilities of different methods in reflecting the magnetic field strength at different heights. In addition, by analyzing the magnetic field orientation through the inclination and azimuthal angles as a function of height, we can assess the models' proficiency in capturing the geometric properties of the field. These metrics collectively demonstrate the methods' competencies not only in emulating the MHD model outputs but also in accurately rendering the comprehensive physics quantities and magnetic field orientation at various atmospheric levels, including the surface and coronal height.

Fig.~\ref{2dhist} shows the 2D histograms that compare predictions from different deep learning methods (GL-FNO, ViT, CNN-RNN, and CNN-LSTM) to the ground truth across various height = 2.0 Mm, 7.0 Mm, and 11.0 Mm. Each method demonstrates a scattered distribution of points along the diagonal, indicating a linear correlation between predicted and true magnetic field strengths. At a height of 2.0 Mm, closer to the surface, the GL-FNO and ViT show a denser clustering of points around the diagonal, a pattern less concentrated in the CNN-LSTM and CNN-RNN. At greater heights of 7.0 Mm and 11.0 Mm, GL-FNO consistently keeps points near the diagonal, showcasing sustained prediction accuracy. On the contrary, ViT, CNN-LSTM and CNN-RNN exhibit a wider spread of points, hinting at a decline in predictive precision.

\begin{figure}[ht]
	\centering
	\includegraphics[width=\linewidth]{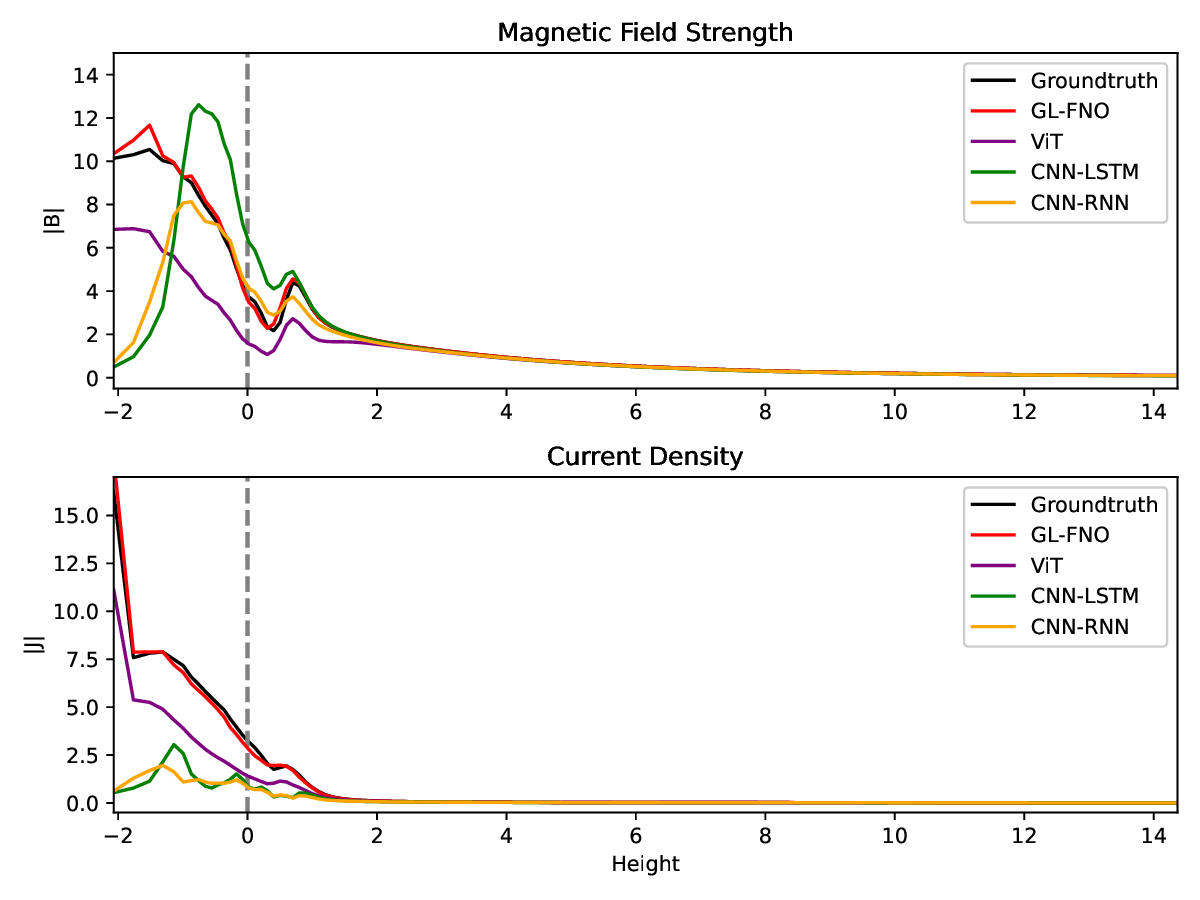}	
	\caption{Comparison of physics quantities derived from GL-FNO, ViT, CNN-RNN, and CNN-LSTM. Upper panel: Variation of magnetic field strength as a function of height. Lower panel: Variation of current density as a function of height.}  
	\label{magnetic_field}       
\end{figure}

Moreover, we have evaluated physics-based measurements varying with height to assess each model's ability to reproduce the MHD model outputs. Fig.~\ref{magnetic_field} displays magnetic field strength ($|B|$) and current density ($|J|$) revealing that, above a height of 2 Mm, all models produce patterns that resemble the ground truth, indicative of a stable environment with low magnetic field strength. Below this height, only the GL-FNO model's predictions align closely with the actual measurements. Similarly, with current density, GL-FNO maintains a near-ground truth trend below a height of 1.5 Mm. At a height above 1.5 Mm, characterized by low current, all deep learning methods follow a similar trajectory. Overall, GL-FNO delivers the most consistent and accurate predictions of these physical quantities, particularly close to the surface where the magnetic settings are complex.

\begin{figure}[ht]
	\centering
	\includegraphics[width=\linewidth]{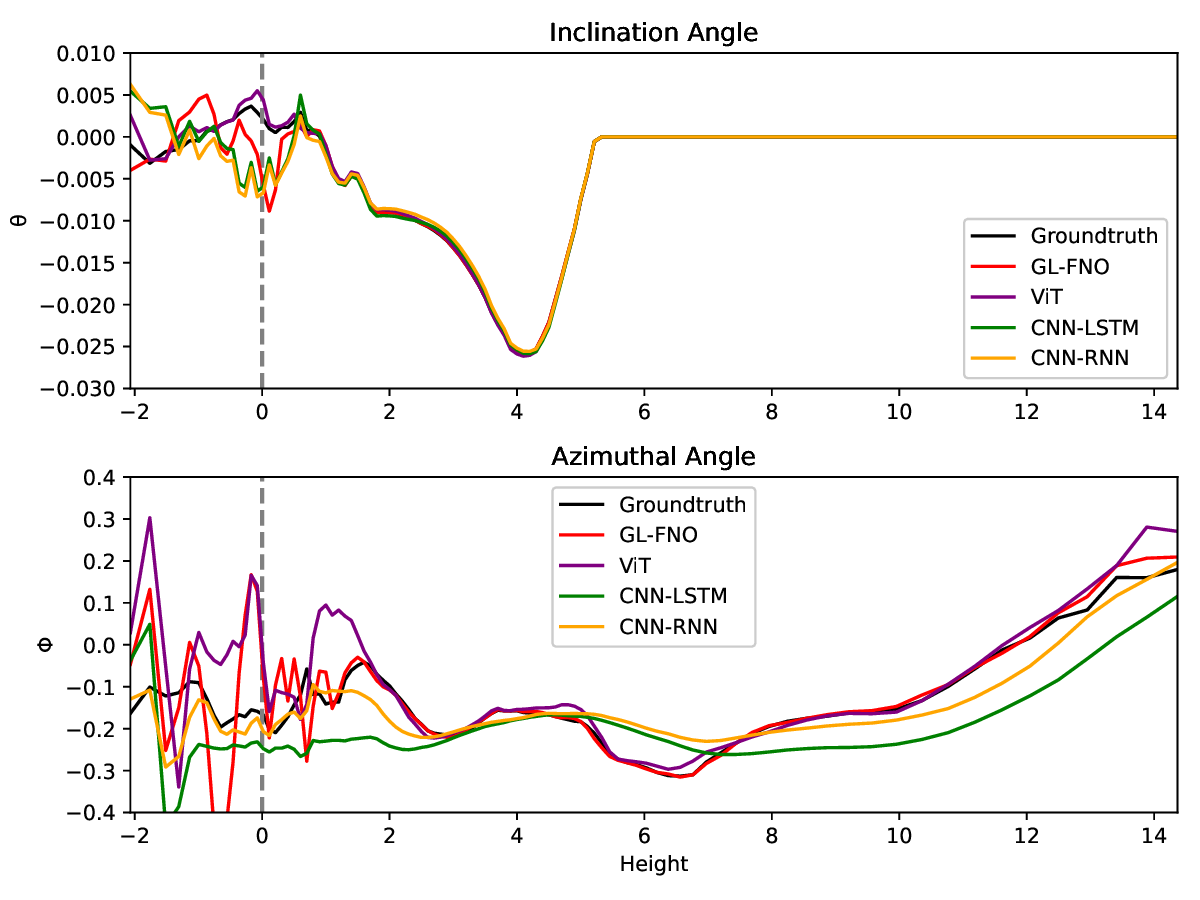}	
	\caption{Comparison of field orientation derived from GL-FNO, ViT, CNN-RNN, and CNN-LSTM. Upper panel: Variation of inclination angle as a function of height. Lower panel: Variation of azimuthal angle as a function of height.}  
	\label{inc_azi_angle}       
\end{figure}

Fig.~\ref{inc_azi_angle} illustrates the variation of the horizontally averaged inclination angle $\theta$ and azimuthal angle $\phi$ as a function of height, derived from different deep learning methods compared to the ground truth. Please note that a binary map has been applied to focus solely on magnetic field strengths that exceed 2.5 times the background level. This approach helps to disregard weaker fields, which tend to be noisier and less significant, while still taking into account their potential influence on the calculations of field orientation metrics. All deep learning methods can reproduce the inclination angles well after height $>$ 1 Mm. At a lower height below 1 Mm, ViT outperforms other models regarding the inclination angles. The azimuthal angle variation, however, demonstrates a significant divergence between the deep learning methods and the ground truth at the lower height (below height = 2 Mm). GL-FNO and ViT appear to align with the ground truth, but GL-FNO converges faster than ViT. Although CNN-LSTM and CNN-RNN follow the general trend,  they do not closely align with the ground truth. At the higher layer (height $>$ 13 Mm), GL-FNO continues to track the ground truth, whereas ViT starts to diverge. The inclination and azimuthal angles are important for understanding the orientation of the magnetic field. The consistency of the GL-FNO with the ground truth suggests it may provide a more physically accurate magnetic field orientation with height, compared with other deep learning methods. 

\section{Conclusion}\label{sec: conclusion}
This paper explores the use of neural operator architectures to accelerate the speed of computational modeling of the solar coronal magnetic field. Specifically, In our proposed GL-FNO, the global branch captures the overall structure and dynamics while the local branch focuses on fine-grained features. This fusion enables our GL-NFO to generalize better across different physical conditions while retaining accuracy in areas requiring detailed resolution. GL-FNO has demonstrated the capability to simulate the complex magnetic solar outer atmosphere compared to state-of-the-art deep learning methods such as FNO, U-NO, U-FNO, ViT, CNN-RNN, and LSTM. The findings from the comparative study reveal that GL-FNO achieves superior prediction accuracy as evidenced by the best test performance among all methods. Compared to traditional physics-based models like MHD, GL-FNO only takes a few seconds to generate the 3D cube while MHD takes days to simulate. 

From the physics perspective, GL-FNO stands out for its precision in predicting the magnetic field topologies, crucially maintaining this accuracy across varying atmospheric heights. GL-FNO exhibits a remarkable adherence to ground truth, particularly below a height of 2 Mm, a region characterized by complex magnetic interactions. Its performance is consistently superior to that of other methods, such as the ViT, CNN-RNN, and CNN-LSTM, especially in critical lower atmospheric layers where precise modeling is most needed. Additionally, GL-FNO's alignment with ground truth in the calculation of field orientation metrics further solidifies its capability to capture the nuanced dynamics of solar phenomena.

\bibliographystyle{IEEEtran}
\bibliography{IEEEfull,Example}

\vspace{12pt}
\appendix
\section*{Experiments with Different Neural Operators} \label{appendix}
In this appendix, we aim to compare the performance of our proposed method with existing advanced neural operators such as FNO~\cite{Li2020FourierNO}, U-shaped Neural operator (U-NO) \cite{rahman2022u}, and U-FNO~\cite{wen2022u}. They are selected for comparison since they are outstanding models in the neural operator literature \cite{azizzadenesheli2024neural}.
\subsection{Neural Operator Model Structures}
In our experiments, FNO, U-NO, and U-FNO have the following model structure. 
\subsubsection{Fourier Neural Operator} FNO follows the structure of the local branch in our proposed GL-FNO structure as shown in Section~\ref{subsubsec: GL-NFO}.

\subsubsection{U-shaped Neural operator}
The U-NO used in our paper is the same structure as described in \cite{rahman2022u}. It employs a U-Net architecture where each layer is composed of FNO blocks with specific configurations. The model processes a single-channel input and predicts 99 output channels, using a hidden dimension of 128. The U-Net structure contains layers with output channels set to $(32, 64, 64, 32)$ and Fourier modes of $((64, 64), (64, 64), (64, 64), (64, 64))$, alongside scaling factors $((1.0, 1.0), (0.5, 0.5), (1, 1), (2, 2))$, which control the spatial resolution at different stages of the network.

\subsubsection{U-FNO}
The U-FNO \cite{wen2022u} used in our paper is a modified version of the origin model architecture. We modified the original 3D Fourier neural network into a 2D Fourier neural network and adjusted the input and output channels to fit our case, setting them to 1 and 99, respectively. We used $(64,64)$ as the number of Fourier modes and 128 as the number of hidden channels. Other aspects remain consistent with the original model. The model includes four Fourier blocks: the first two Fourier blocks follow the structure in \cite{li2020neural}, while the last two Fourier blocks incorporate a U-net structure to help merge multi-scale features, further enhancing the output.

\begin{table}[!htbp]
    \caption{Test performance comparison of GL-FNO, FNO, U-NO, and U-FNO for $\textbf{B}$ in terms of MSE, $R^2$, RE, MAE, PSNR, and SSIM. }
    \label{table2}       
    {\scriptsize
    \centering
    \begin{tabular}{ccccccc}
        \toprule
        Model &  MSE $\downarrow$& $R^2 	\uparrow$ & RE $\downarrow$& MAE $\downarrow$ & PSNR $\uparrow$ & SSIM $\uparrow$\\ 
        \midrule
        \textbf{GL-FNO} &  \textbf{0.0331} & \textbf{0.9668} & \textbf{0.1756} & \textbf{0.0672}  & \textbf{50.58} & \textbf{0.9883} \\ 
        FNO &  0.0502 & 0.9498 & 0.2205 & 0.0831  & 43.12 & 0.9227 \\ 
        U-NO & 0.0914  & 0.9086 & 0.2931 & 0.1105  & 46.08  & 0.9697 \\ 
        U-FNO  & 0.0564 & 0.9435 & 0.2314 & 0.0967 & 48.07 & 0.9794\\ 
        \bottomrule
    \end{tabular}
    }
\end{table}
\subsection{Test Performance with Different Neural Operators}
Table~\ref{table2} summarizes the performance of GL-FNO, FNO, U-NO, and U-FNO in terms of MSE, $R^2$, RE, MAE, PSNR, and SSIM. Together with Table~\ref{table1}, neural operators all perform much better than ViT, CNN-RNN, and CNN-LSTM. Specifically, GL-FNO, FNO, U-NO, and U-FNO can achieve $R^2>0.9$. This further verifies that neural operators are a good choice for scientific computing applications. Among neural operator methods, GL-FNO is the most effective one.
\end{document}